\documentclass[prl,twocolumn,nofootinbib]{revtex4-1}
\usepackage{graphicx}
\usepackage{color}
\usepackage{dcolumn}
\usepackage{bm}
\usepackage{amssymb}
\usepackage[bottom]{footmisc}
\usepackage{footnote}
\usepackage{lipsum}
\usepackage{wrapfig}
\usepackage{sidecap}
\renewcommand{\thefootnote}{\fnsymbol{footnote}}

\newcommand\blfootnote[1]{%
\begingroup
\renewcommand\thefootnote{}\footnote{#1}%
\addtocounter{footnote}{-1}%
\endgroup}
\begin{document}


\title{Room temperature 2D ferromagnetism in few-layered 1$T$-CrTe$_{2}$}

\author{Xingdan Sun,$^{1,2,*}$ Wanying Li,$^{1,2,*}$ Xiao Wang,$^{3,4,5,*}$ Qi Sui,$^{6,*}$ Tongyao Zhang,$^{7,8}$ Zhi Wang,$^{1,2}$ Long Liu,$^{1,2}$ Da Li,$^{1,2}$ Shun Feng,$^{1,2,11}$ Siyu Zhong,$^{9}$ Hanwen Wang,$^{1,2}$ Vincent Bouchiat,$^{10}$ Manuel Nunez Regueiro,$^{10}$ Nicolas Rougemaille,$^{10}$ Johann Coraux,$^{10}$ Zhenhua Wang,$^{1,2}$ Baojuan Dong,$^{7,8}$ Xing Wu,$^{9}$ Teng Yang,$^{1,2\dagger}$ Guoqiang Yu,$^{3,4,5\dagger}$ Bingwu Wang,$^{6\dagger}$ Zheng Vitto Han$^{1,2,7\dagger}$ Xiufeng Han,$^{3,4,5}$ Zhidong Zhang$^{1,2}$}

\affiliation{$^{1}$Shenyang National Laboratory for Materials Science, Institute of Metal Research, Chinese Academy of Sciences, Shenyang 110016, China}
\affiliation{$^{2}$School of Material Science and Engineering, University of Science and Technology of China, Anhui 230026, China}
\affiliation{$^{3}$Beijing National Laboratory for Condensed Matter Physics, Institute of Physics, Chinese Academy of Sciences, Beijing 100190, China}
\affiliation{$^{4}$Center of Materials Science and Optoelectronics Engineering, University of Chinese Academy of Sciences, Beijing 100049, China}
\affiliation{$^{5}$Songshan Lake Materials Laboratory, Dongguan, Guangdong 523808, China}

\affiliation{$^{6}$Beijing National Laboratory for Molecular Sciences, Beijing Key Laboratory of Magnetoelectric Materials and Devices, College of Chemistry and Molecular Engineering, Peking University, Beijing, 100871 China}
\affiliation{$^{7}$Collaborative Innovation Center of Extreme Optics, Shanxi University, Taiyuan 030006, P.R.China}
\affiliation{$^{8}$State Key Laboratory of Quantum Optics and Quantum Optics Devices, Institute of Opto-Electronics, Shanxi University, Taiyuan 030006, P. R. China}
\affiliation{$^{9}$Shanghai Key Laboratory of Multidimensional Information Processing Department of Electrical Engineering East China Normal University, 500 Dongchuan Road, Shanghai 200241, China}
\affiliation{$^{10}$University of Grenoble Alpes, CNRS, Institut N\'eel, F-38000 Grenoble, France.}
\affiliation{$^{11}$School of physical science and technology, ShanghaiTech University, Shanghai 200031, China.}


\maketitle
\blfootnote{\textup{*} These authors contribute equally.}

\blfootnote{$^\dagger$Corresponding to: yangteng@imr.ac.cn, wangbw@pku.edu.cn, guoqiangyu@iphy.ac.cn, and vitto.han@gmail.com}

\textbf{Spin-related electronics using two dimensional (2D) van der Waals (vdW) materials as a platform are believed to hold great promise for revolutionizing the next generation spintronics. Although many emerging new phenomena have been unravelled in 2D electronic systems with spin long-range orderings, the scarcely reported room temperature magnetic vdW material has thus far hindered the related applications. Here, we show that intrinsic ferromagnetically aligned spin polarization can hold up to 316 K in a metallic phase of 1$T$-CrTe$_{2}$ in the few-layer limit. This room temperature 2D long range spin interaction may be beneficial from an itinerant enhancement. Spin transport measurements indicate an in-plane room temperature negative anisotropic magnetoresistance (AMR) in few-layered CrTe$_{2}$, but a sign change in the AMR at lower temperature, with -0.6$\%$ at 300 K and +5$\%$ at 10 K, respectively. This behavior may originate from the specific spin polarized band structure of CrTe$_{2}$. Our findings provide insights into magnetism in few-layered CrTe$_{2}$, suggesting potential for future room temperature spintronic applications of such 2D vdW magnets.}


 \bigskip
 \bigskip

Spin-polarized 2D vdW layers have been a cutting-edge topic in recent years, as it not only provides opportunities for new-concept nanostructures,\cite{CrI32017, CGT2017, WangZhi2018, JiangShengwei2018, Yuanbo_Nature2018} but also serves as a test-bed for novel spin physics.\cite{WuShiwei2019nature, GaoChunlei_arXiv2019} Especially, spintronic devices built with vdW magnets have aroused great interest.\cite{Xiaoxi} For example, magnetoresistance in spin-filtered magnetic  vdW heterojunctions was manifested to be as high as 10000$\%$, far superior to that of conventionally grown magnetic thin films.\cite{Song1214} More recently, current-induced magnetic switch was succeeded in the Fe$_{3}$GeTe$_{2}$ few-layers,\cite{YuGuoqiang2019} suggesting vdW magnet a versatile platform for nanoelctronics. However, most of those demonstrated vdW spintronic devices are functioning at low temperature, which are practically not preferable.

   \begin{figure*}[ht!]
   \includegraphics[width=0.85\linewidth]{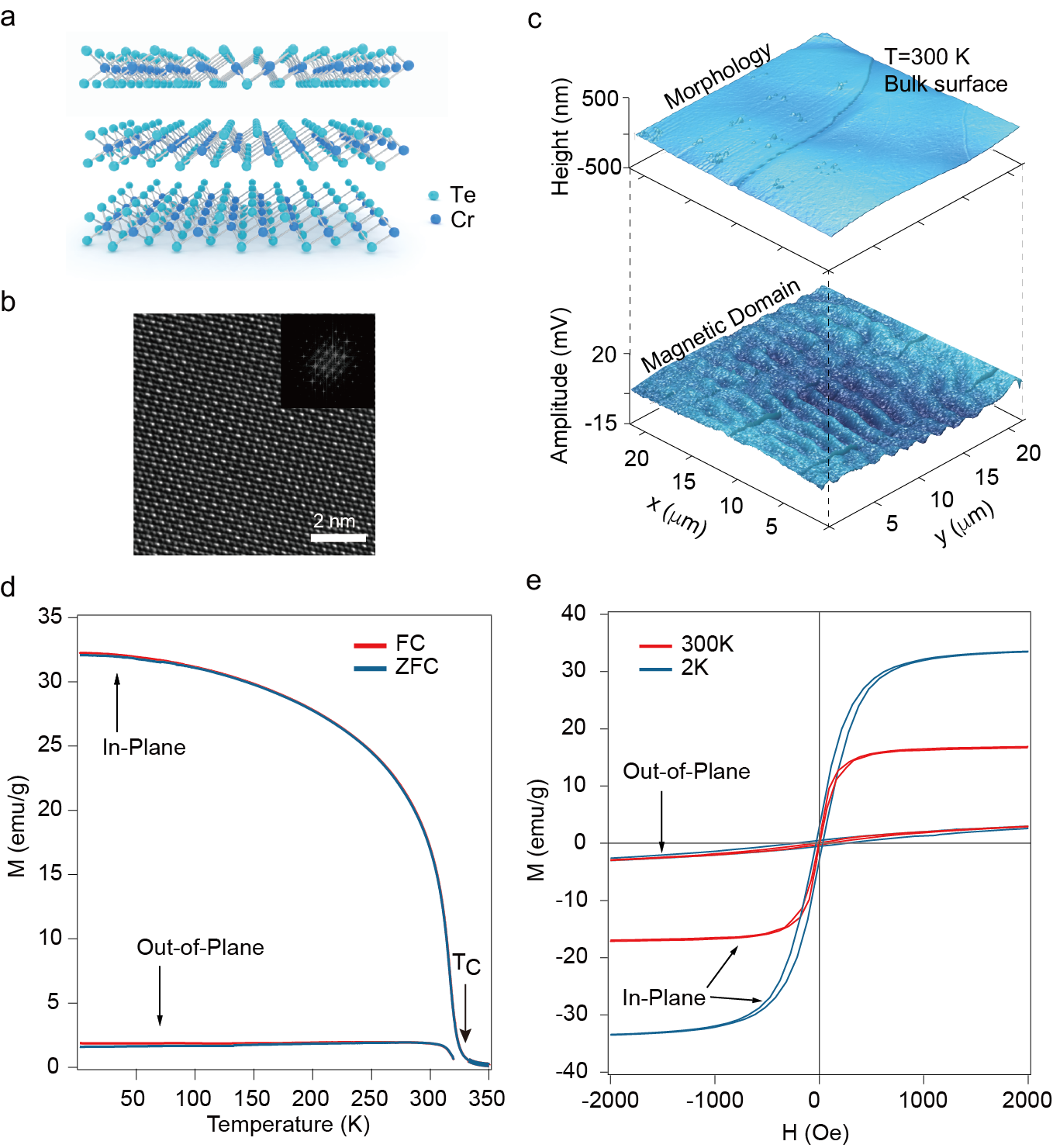}
   \caption{\textbf{Characterizations of bulk 1$T$-CrTe$_{2}$.} (a) Schematics of CrTe$_{2}$ crystal structure. (b) Transmission electron micrography of CrTe$_{2}$ crystal. Inset shows the fast Fourier transform pattern of the obtained atomic micrograph. (c) AFM morphology image of a typical CrTe$_{2}$ surface freshly exfoliated from a bulk crystal, with its amplitude profile plotted in lower panel, indicating a fringe-like magnetic domain. Images are obtained at room temperature. (d) ZFC-FC curves of bulk CrTe$_{2}$. Solid arrow indicates the Curie temperature. (e) Magnetization hysteresis loops of bulk CrTe$_{2}$ at 2 K (blue lines) and 300 K (red lines), respectively.
   }
   \end{figure*}

Interestingly, even in those vdW magnets with relatively low intrinsic Curie temperature (T$_{C}$) such as Fe$_{3}$GeTe$_{2}$, room temperature 2D ferromagnetism can often be realized via certain extrinsic manners, such as strong electron doping with an ionic gate,\cite{Yuanbo_Nature2018} or the enhancement of magnetic anisotropic energy through a micrometer-sized geometric confinement.\cite{QiuZiqiang_NanoLett_2018} Nevertheless, these methods are in principle incompatible with large scale solid state device applications. Demonstrations of spintronic devices based on intrinsic room temperature ferromagnetic vdW materials remain challenging.

   \begin{figure*}[ht!]
   \includegraphics[width=0.85\linewidth]{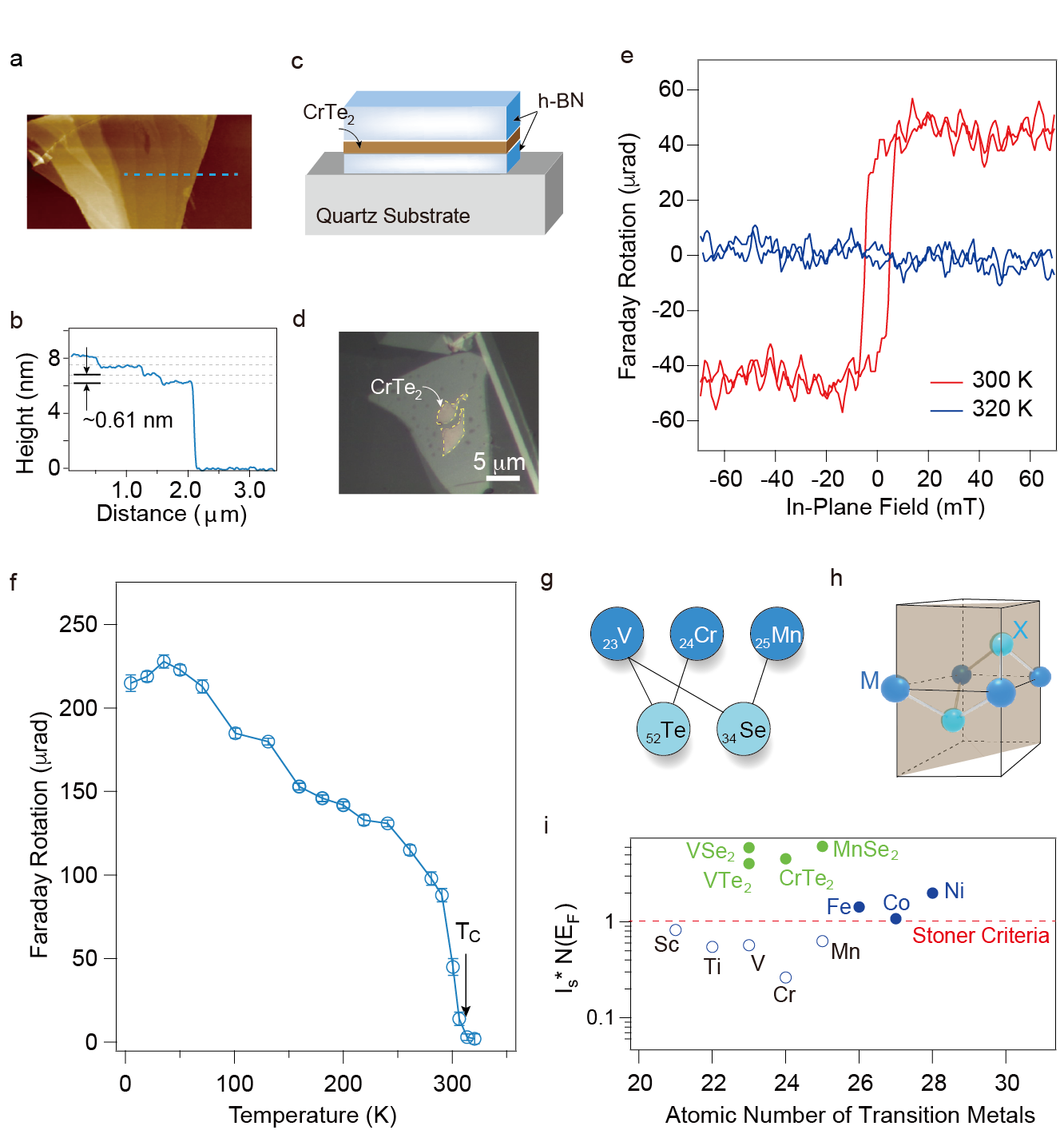}
   \caption{\textbf{Magnetic properties of few-layered CrTe$_{2}$.} (a) AFM morphology image of a typical few-layered CrTe$_{2}$. (b) Height profile along the dashed line in (a), with the layer thickness measured to be around 0.61 nm, as indicated by the arrows. (c) Schematic picture and (d) optical image of BN-encapsulated few-layered CrTe$_{2}$ deposited on a quartz substrate for optical measurements. (e) Faraday rotation of a typical 10 nm CrTe$_{2}$ flake at 300 K and 320 K, respectively. (f) M-T curve extracted from the Faraday rotation at 40 mT. (g) A summarization of the four types of known room temperature vdW ferromagnets. (h) Unit cell of MX$_{2}$. (i) Analysis of potential band ferromagnetism of  several materials using the Stoner criteria.}

   \label{fig:fig2}
   \end{figure*}

To date, intrinsic vdW ferromagnets with T$_{C}$ above 300 K have been reported in a very limited library of materials, including epitaxially or chemical vapour/exfoliation synthesized 2D films of MnSe$_{x}$\cite{MnSe}, VTe$_{2}$,\cite{VTe2} and VSe$_{2}$\cite{KianPingLoh_AM2019, Bonila_NN2017}. Yet it is still plausible about their origin of magnetism since MnSe$_{x}$ and VTe$_{2}$ were reported in absence of an experimentally well defined T$_{C}$.\cite{MnSe, VTe2} The origin of ferromagnetism in the 2D limit in MnSe$_{x}$ and VSe$_{2}$ are yet to be fully understood, as bulk forms of them are known to be non-magnetic.\cite{MnSe2-bulk-pm1, MnSe2-bulk-pm2, VSe2-bulk-pm1, VSe2-bulk-pm2} Recent theoretical and experimental analysis indicate that a charge density wave phase may be playing key roles in the magnetic order of VSe$_{2}$.\cite{GaoHongjun_SB2018,PRB_241404_2019, PRL_196402_2018,BaoLihong_NanoLett_2019} During the preparation of this manuscript, we notice that bulk vdW Fe$_{5}$GeTe$_{2}$ has a T$_{C}$ of about 310 K, while its few-layer samples has T$_{C}$ of about 270 K, slightly lower than room temperature.\cite{Fe5GeTe2}

It is noticed that Cr$_{x}$Te$_{y}$ has been one of the overlooked families so far in the hunt for room temperature vdW magnets. Among them, 1$T$-CrTe$_{2}$ is a layered compound with ferromagnetism critical temperature of $\sim$320 K in its bulk phase, making it a candidate for the investigation of 2D ferromagnetism.\cite{CrTe2_bulk_Tc_2015} In this work, we performed mechanical exfoliation of a bulk vdW crystal of 1$T$-CrTe$_{2}$, and found that the Curie temperature above 300 K for bulk CrTe$_{2}$ can be retained down to the few-layered 2D limit. Our simulations suggest that exchange coupling due to an enhancement of itinerant type was the source of room temperature ferromagnetism in both bulk and few-layered CrTe$_{2}$. The 2D ferromagnetism in few-layered CrTe$_{2}$ was further investigated in the framework of in-plane AMR at different temperatures. Our findings demonstrate that few-layered CrTe$_{2}$ can be a candidate for next generation room temperature spin-related 2D nanoelectronics.

\section{Results}
\textbf{Characterizations of bulk phase of CrTe$_{2}$.} Bulk CrTe$_{2}$ single crystals were prepared (see Methods) and were confirmed via x-ray diffraction (Supplementary Figure 1). It is known that 1$T$-CrTe$_{2}$ has a layered structure with lattice symmetry of space group $P\overline{3}m1$ (Fig. 1a). To confirm the crystalline structure of the as-synthesized crystals, we performed transmission electron micrograph (TEM) of an exfoliated CrTe$_{2}$ flake, as shown in Fig. 1b. Clear hexagonal lattice structure, resembling the top view of Fig. 1a, can be seen. Atomic force microscopy was used to characterize the morphology (upper panel of Fig.1c) with its corresponding magnetic domain shown in the lower panel of Fig. 1c. As can be seen, at room temperature, the bulk CrTe$_{2}$ has a fringe-like magnetic domain, with the domain width of about 1 $\sim \mu$m. Zero-field-cooled (ZFC) and field-cooled (FC) thermal magnetization curves of  bulk CrTe$_{2}$ were obtained. As shown in Fig. 1d, the Curie temperature T$_{C}$ is determined to be around 320 K with a 1,000 Gauss magnetic field applied. To clarify the magnetization behavior of bulk CrTe$_{2}$, M-H loops were examined at 10 K and 300 K, which are both typically ferromagnetic, as shown in Fig. 1e.

\bigskip

\textbf{Magnetic properties of exfoliated few-layered CrTe$_{2}$.} In order to study the magnetism of CrTe$_{2}$ in the 2D limit, we carried out mechanical exfoliation using the standard scotch tape method.\cite{graphene_2004} Atomic Force Microscope (AFM) scan of a typical CrTe$_{2}$ flake (thinnest part of $\sim$6.1 nm) is shown in Fig. 2a-b. It is found that the yield of atomically thin CrTe$_{2}$ is rather low compared to those easily exfoliated layered compound such as graphene, and the thinnest flake we could obtain was $\sim$3 nm (Supplementary Figure 2). Moreover, similar to other telluride materials, few-layered CrTe$_{2}$ in air is easily degraded, leading to failure for any further characterizations (Supplementary Fig. 3). Hexagonal boron nitride (h-BN) encapsulation in a glove box therefore can be a solution to avoid air-degradation of ultra-thin CrTe$_{2}$ flakes. A detailed Raman spectra measurements of few-layered CrTe$_{2}$ is given in Supplementary Figures 4-6. It is found that intrinsic thin flakes of CrTe$_{2}$ show Raman shift peaks at about 99.7 and 132 cm$^{-1}$, while degraded ones will shift its characteristic peaks to 120 and 140 cm$^{-1}$ with higher intensity. It thus provides us a quick way to select the pristine thin flakes for magnetic and electrical measurements.

With protection from air, the h-BN sandwiched thin layers of CrTe$_{2}$ were deposited onto transparent quartz substrates for Faraday measurements, as sketched in Fig. 2c. Oblique incident probe beam with respect to the sample plane was applied to monitor the in-plane component of the magnetization in the few-layered CrTe$_{2}$ flakes (Supplementary Figure 7). Spot size of about 2 $\mu$m diameter laser with wavelength of 800 nm was used, and an optical image of typical sample is shown in Fig. 2d. A typical 10 nm CrTe$_{2}$ sample was loaded in a vacuum chamber and measured in a transmission configuration  (Supplementary Figure 7) with its magnetization hysteresis loops (M-H loops). As shown in Fig. 2e, the few-layered CrTe$_{2}$ exhibits characteristic ferromagnetic M-H loops at 300 K, while the loop became a linear background above 320 K. We extracted the saturated Faraday effect signal at H=40 mT, and thus obtained an effective M-T curve, shown in Fig. 2f. The general trend of M-T in few-layered CrTe$_{2}$ is rather similar to that measured in its bulk, with a T$_{C}$ to be around 316 K in few-layered CrTe$_{2}$, as indicated by the solid arrow in Fig. 2f.

\begin{figure*}[ht!]
\includegraphics[width=0.85\linewidth]{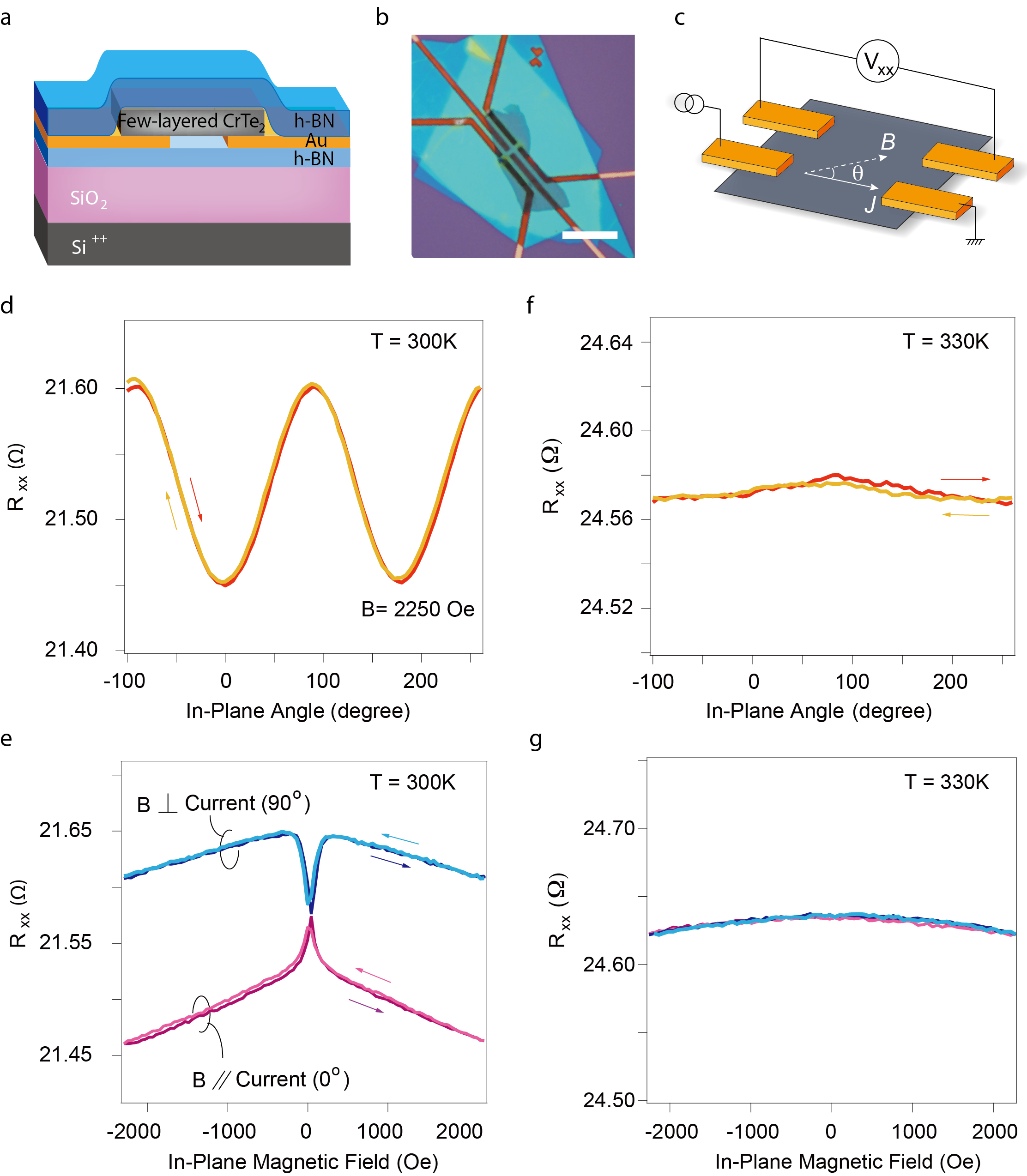}
\caption{\textbf{Anisotropic magnetoresistance of few-layered CrTe$_{2}$ at 300 K.} (a) Art view and (b) Optical image of a typical h-BN/CrTe$_{2}$/h-BN device. Scale bar in (b) is 10 $\mu$m. (c) Illustration of the measurement of in-plane magnetic field with an angle $\theta$ against the current $\textbf{\textit{J}}$. (d) R$_{xx}$ of few-layered CrTe$_{2}$ as a function of $\theta$ with an 2250 Oe magnetic field applied. (e) R$_{xx}$ of few-layered CrTe$_{2}$ as a function of magnetic field $\textbf{\textit{B}}$, with $\theta$=0$^{o}$ (pink lines) and 90$^{o}$ (blue lines), respectively. (f) and (g) are R$_{xx}$ vs $\theta$, and R$_{xx}$ vs $\textbf{\textit{B}}$ with the same configurations as in (d)-(e) measured at 330 K.}

\label{fig:fig3}
\end{figure*}

\begin{figure*}[ht!]
\includegraphics[width=1\linewidth]{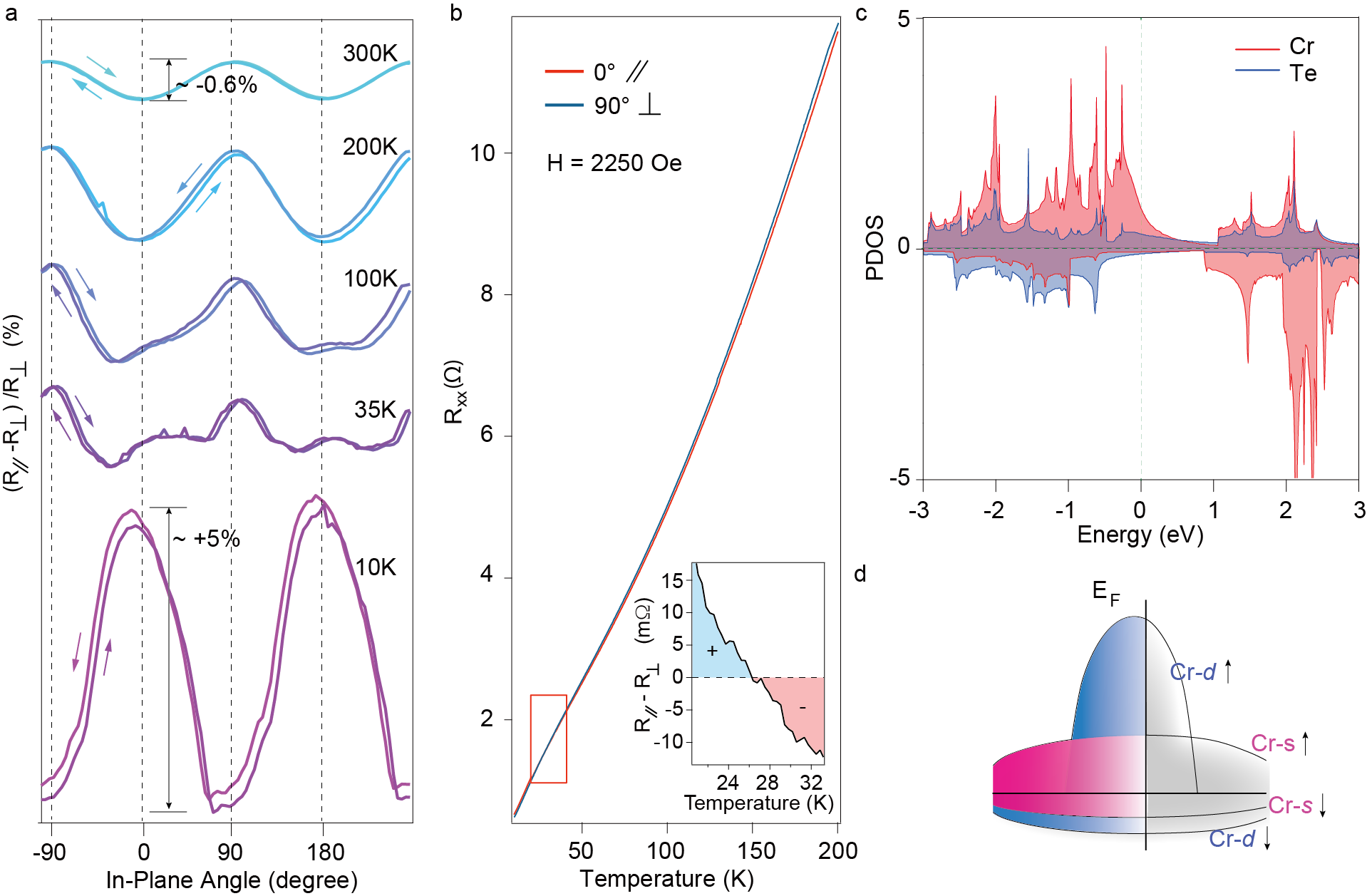}
\caption{\textbf{Mechanism of AMR in few-layered CrTe$_{2}$.} (a)  $r_{\textrm{AMR}}$ of few-layered CrTe$_{2}$ at different temperatures. (b) R$_{xx}$ as a function of temperature at $\theta$=0$^{o}$ (red line) and 90$^{o}$ (blue line). An in-plane magnetic field of 2250 Oe was applied. Inset shows the difference between R$\parallel$ and R$\perp$, with negative and positive values separated at the temperature of about 26.5 K. (c) Partial DOS (PDOS) from both Cr and Te for spin up and spin down sub bands calculated for monolayer CrTe$_{2}$. (d) Schematic image of the decomposed Cr-s and Cr-d orbitals near the Fermi level.}

\label{fig:fig4}
\end{figure*}

Interestingly, among the known room temperature 2D ferromagnets, i.e., VSe$_{2}$, MnTe$_{2}$, MnSe$_{x}$, and CrTe$_{2}$, they all belong to a family of transitional metal dichalcogenides MX$_{2}$ with octahedral units (1$T$), as illustrated in the cartoon in Fig. 2g-2h. The triangle M sublattice of the 1$T$ major phase gives rise to a first-neighbor coordination number of 6. Besides the indirect superexchange coupling, itinerant exchange~\cite{Stoner36} should be playing an essential role due to the metallic nature of these room temperature vdW ferromagnets. To test whether itinerant (or band) ferromagnetism appears in a metal, one usually refers to the so-called Stoner criteria, namely the I$_s$$*$N(E$_F$) $\geq$ 1 condition~\cite{Mohn03}. The I$_s$ quantity is called Stoner exchange integral, which depends very little on bonding environments~\cite{Mohn03}, and N(E$_F$) is the electronic density of states at the Fermi level. In Fig. 2i, four typical MX$_{2}$ 2D room temperature ferromagnets are given, and compared to both the corresponding elemental metals and ferromagnetic Fe, Co and Ni elemental metals. Expectedly, only Fe, Co and Ni among all the given elemental metals meet the Stoner criteria to exhibit band ferromagnetism. However, the 2D VSe$_{2}$, CrTe$_{2}$, MnTe$_{2}$ and MnSe$_{2}$ have I$_s$$*$N(E$_F$) much increased upon that of the corresponding elementary metals, and the Stoner criteria is then met for a band ferromagnetism in these 2D materials. The chalcogenide atoms are important in boosting ferromagnetism in terms of the following: (i) Double superexchange. Neighbouring magnetic elements M (= V, Cr, Mn) are coupled to each other by superexchange through M-X-M bonds of nearly 90$^{o}$ bond angle. (ii) Enhancement of N(E$_F$). Electronic density of states (DOS) at Fermi level of MX$_2$ is enhanced upon N(E$_F$) of the corresponding elemental metals, due to orbital rearrangement by the X atoms. The partial DOS analysis shows that N(E$_F$) in 1$T$-MX$_2$ is overwhelmingly from the orbitals of M atoms (Supplementary Figure 8). (iii) The Slater criterion~\cite{Slater30}. In this criterion, a ratio $\frac{r}{2 r_a}$ of interatomic distance $r$ over the double atomic radius 2$r_a$ has to exceed 1.5 for the occurrence of ferromagnetism~\cite{Slater30,Omar75}, with Fe, Co and Ni elementary metals satisfying this criterion, while V, Cr and Mn not. The existence of chalcogenide X atoms in those 2D 1$T$-MX$_2$ may play a role in enlarging the interatomic separation of V, Cr and Mn, so as to meet the Slater criterion.

To further quantify the strength of itinerant exchange interaction, mean-field solution to the Heisenberg model is employed~\cite{Blundell01,Mohn03}, namely, T$_c$ = $\frac{2zJS(S+1)}{3 k_B}$, in which $z$ is the coordination number (for instance $z$ = 6 in 1$T$ MX$_2$), $J$ is the exchange, and $S$ is the spin angular momentum (orbital angular momentum $L$ = 0 due to orbital quenching in octahedral MX$_2$). We therefore evaluate the exchange $J$ in terms of $J = \frac{3 k_B T_c}{2zS(S+1)}$ and show it in Supplementary Figure 9. The exchange strength $J$, around 70-120 $meV$, is comparable to that in the  elemental ferromagnets and 2D itinerant ferromagnets, but much larger than that in non-itinerant 2D magnets such as CrI$_3$ (Supplementary Figure 9). We have to admit that the evaluation of exchange is more or less simplified due to the lack of consideration of several factors including magnetocrystalline anisotropy, and the exceptional cases of charge density wave (CDW) in VSe$_{2}$ and VTe$_{2}$ \cite{Sugawara19,yu19}. A thorough understanding on the room temperature ferromagnetism in the 2D 1$T$-MX$_2$ remains an open question.

\section{Discussion}

\textbf{Spin transport of CrTe$_{2}$ in the 2D limit.} We now come to the investigation of spin transport behaviour in few-layered CrTe$_{2}$. Different ways of protection methods were performed, including Pt thin film capping and BN-encapsulation (Supplementary Figures 10-11), in order to enhance the air-stability of few-layered CrTe$_{2}$. In general, ultra thin CrTe$_{2}$ can be quite stable as long as heating is well avoided during the whole fabrication process. As sketched in Fig. 3a, the CrTe$_{2}$ flakes were encapsulated by two h-BN flakes, with Au electrodes patterned in between via lithography. Optical image of a typical BN-encapsulated device is shown in Fig. 3b. Because of the in-plane magnetic anisotropy, no detectable Hall signal was obtained in those devices. In the following, we mainly focus on the longitudinal resistance R$_{xx}$, with the configuration for measurements illustrated in Fig. 3c.

Fig. 3d shows R$_{xx}$ at 2250 Oe, recorded in a typical CrTe$_{2}$ device with thickness of 10 nm, as a function of the angle $\theta$ between the in-plane magnetic field $\textbf{\textit{B}}$ and the electrical current $\textbf{\textit{J}}$ flowing between source and drain electrodes. A two-fold cos(2$\theta$)-like oscillation behaviour was observed, similar to previously reported.\cite{Takata_2017, Tsunoda_2009} Fig. 3e shows the corresponding magnetoresistance (M-H curves) at 300 K for $\theta$=0$^{o}$, and 90$^{o}$, respectively (data at more temperatures shown in Supplementary Figure 12). It is noted that when heated up to 330 K, the AMR related features disappear, and can be re-appearing once the sample is cooled back down to 300 K. This is a prove that the ferromagnetic-paramagnetic phase transition in few-layered CrTe$_{2}$ is reversible. This observed negative in-plane AMR behaviuor (reproduced in multiple samples as shown in the Supplementary Figures 13) is seen in some conventional half-metallic ferromagnets such as Ni$_{3}$FeN \cite{Takata_2017}, as well as in recently reported  Mn-doped Bi$_{2}$Se$_{3}$ \cite{Mn_Bi2Se3_AMR}. On the contrary, many other ferromagnets, such as Ni and Fe,\cite{Fe_Co_Ni_AMR} and vdW materials  Fe$_{3}$GeTe$_{2}$\cite{AMR_FGT_arXiv2019} (with much smaller $r_{\textrm{AMR}}$) have positive in-plane AMR.

When further cooled down, as shown in Fig. 4a, the shape of $r_{\textrm{AMR}}$ became more or less deviated from the cos(2$\theta$) symmetry. At 10 K, the general  $r_{\textrm{AMR}}$ changes its sign from negative into positive, with its amplitude reaching 5$\%$, at the same order of magnitude as compared the maximum AMR reported in the NiFe alloys \cite{Ni3Fe_AMR, Ni3Fe_AMR1}. However, NiFe alloys such as Ni$_{3}$FeN thin film has positive  $r_{\textrm{AMR}}$ at room temperature and negative  $r_{\textrm{AMR}}$ at low temperature \cite{Takata_2017}, which are quite opposite with respect to that observed in few-layered CrTe$_{2}$. According to the spin scattering model \cite{Kokado_model} $r_\textrm{AMR}=\frac{R_{\parallel}-R_{\perp} }{R_{\perp}}\propto -( N_{\uparrow}^{d}-N_{\downarrow}^{d})\cdot (\sigma_{\uparrow}- \sigma_{\downarrow})$, the competition between conducting electrons at each spin branch will be the origin of the sign change on the experimentally obtained  $r_{\textrm{AMR}}$ at different temperatures. 

We recorded the R-T curves of R$_{xx}$ for the same device measured in Fig. 4a, at $\theta$=0$^{o}$ (red solid line) and 90$^{o}$ (blue solid line), with a fixed in-plane magnetic field of 2250 Oe, as shown in Fig. 4b. A crossing point can be seen in the boxed area, as also zoomed in the inset, at the temperature of about 26.5 K. By computing the partial density of states (PDOS) of mono-layer CrTe$_{2}$, it is clearly seen in Fig. 4c and Supplementary Figure 14 that $N_{\uparrow}^{d}$ is much larger than $N_{\downarrow}^{d}$. Nevertheless, as illustrated in Fig. 4d (more details refer to Supplementary Figure 14), at the Fermi level, the spin majority ($\uparrow$) Cr $d$ orbital is more localized than the spin minority ($\downarrow$) Cr $d$ orbital, meanwhile, Cr $s$ orbitals for both spin majority and minority are delocalized. It is therefore expected that spin up electrons contribute to the minority of conductivity, due to a much larger magnitude of $s$-$d$ scattering in spin up channel than in spin down channel, making $\sigma_\uparrow-\sigma_\downarrow <$ 0, thus a positive  $r_{\textrm{AMR}}$ at low temperature. When thermal excitation takes over with increasing temperature, more spin up Cr $d$ orbitals are occupied, suppressing $s$-$d$ scattering and giving rise to a sign change in the $\sigma_\uparrow-\sigma_\downarrow$ term, thus a negative  $r_{\textrm{AMR}}$ up to 300 K.  

\bigskip

In conclusion, after the three reported room temperature 2D ferromagnets, we have found the 4$^{th}$ one, i.e., CrTe$_{2}$. By optical and electrical measurements, ferromagnetism was proven to be prevailing in CrTe$_{2}$ in the few-layered limit, with Cuire temperature above 300 K. Detailed spin transport measurements suggest that a half-metalicity in its spin-polarized band structure gives rise to a -0.6 $\%$ $r_\textrm{AMR}$ at 300 K, which changes into a +5 $\%$ $r_\textrm{AMR}$ at 10 K. Our studies reveal that vdW 2D ferromagnet CrTe$_{2}$ can be a candidate for future room temperature spintronic applications, since it can be, in principle, further implemented into in-plane spin valves, as well as large size flexible spin devices.

\section{Methods}

The 1$T$-CrTe$_{2}$ single crystals were synthesized indirectly by oxidation of KCrTe$_{2}$. The parent compound KCrTe$_{2}$ was prepared by a molar mixture of the corresponding elements (nK:nCr:nTe=1:1:2) under argon atmosphere. The mixture was heated at 900 $^{o}$C for seven days in an evacuated quartz tube and then cooled to room temperature at the rate of 10 $^{o}$C h$^{-1}$. The as-synthesized black parent compound KCrTe$_{2}$ was dispersed in acetonitrile and the excess iodine was added very slowly. After stirring for 12 hours, the residue was washed with acetonitrile for three times and the brilliant dark gray platelets 1$T$-CrTe$_{2}$ were obtained. The h-BN (crystals from HQ Graphene) encapsulated CrTe$_{2}$ devices were fabricated using the dry-transfer methods (Supplementary Figures 31)\cite{WangZhi2018} in a glove box. A Bruker Dimension Icon AFM was used for thicknesses and morphology tests. Raman measurements were performed by an HR 800 JobinYvon Horiba polarized Raman spectroscopy. The electrical performances of the devices were measured using a physical properties measurement system (PPMS, Quantum Design) and a probe station (Cascade Microtech Inc. EPS150) under ambient conditions.

The electronic properties in this work were calculated by using the first-principles density functional theory as implemented in the \textsc{VASP} code\cite{VASP}. The electron-ion interaction and electronic exchange-correlation interaction were respectively described by Projector augmented wave (PAW) pseudopotentials~\cite{PAWPseudo} and the Perdew-Burke-Ernzerhof (PBE)~\cite{PBE} functional. The electronic kinetic energy cutoff for plane-wave basis was set to be $520$~eV. The energy criterion for reaching self-consistency was set to be $10^{-8}$~eV. The Brillouin zones were sampled using the $\Gamma$-centered Monkhorst-Pack mesh by $20{\times}20{\times}1$ k-points and $40{\times}40{\times}1$ k-points for self-consistency and  electronic density of states, respectively.~\cite{Monkhorst-Pack76}

\section{\label{sec:level1}DATA AVAILABILITY}
The data that support the findings of this study are available from the corresponding authors upon reasonable request.

\section{\label{sec:level2}ACKNOWLEDGEMENT}
This work is supported by the National Key R$\&$D Program of China (2017YFA0206302 $\&$ 2017YFA0206200) and the National Natural Science Foundation of China (NSFC) with Grants 11974357, U1932151, and 51627801. G.Y. and X.H. thank the finical supports from the National NSFC with Grant No.11874409. T.Yang acknowledges supports from the Major Program of Aerospace Advanced Manufacturing Technology Research Foundation NSFC and CASC, China (No. U1537204). Z. Han acknowledges the support from the Program of State Key Laboratory of Quantum Optics and Quantum Optics Devices (No. KF201816).

\section{Author contributions}
Z.H., B.W., G.Y., T.Y., and Z.Z. conceived the experiment and supervised the overall project. X.S. and W.L. fabricated the devices; X.W.(Xiao Wang), X.S., W.L., D.L., X.H., and G.Y. carried out measurements of spin transport. S.F. performed Raman measurements. T.Z. carried out Faraday rotation measurements; B.D. and T.Y. conducted the theoretical simulations. B.W., Q.S., B.V., C.J., N.R., and M.R. fabricated bulk crystals and performed characterizations of bulk magnetic properties. S.Z. and X.W.(Xing Wu) carried out TEM experiments. L.L. contributed in Pt sputtering. Z.W. discussed on the manuscript. The manuscript was written by Z.H. and X.S. with discussion and inputs from all authors.

\section{ADDITIONAL INFORMATION}
Competing interests: The authors declare no competing financial interests.

\end{document}